\documentclass[twocolumn,10pt,aps,pre,notitlepage,superscriptaddress,
longbibliography]{revtex4-1}

\usepackage{graphicx}
\usepackage{dcolumn}
\usepackage[english]{babel}
\usepackage{bm}
\usepackage{color}
\usepackage{amsmath}
\usepackage{amssymb}
\usepackage{ulem}
\usepackage{bm}
\usepackage{float}
\usepackage{mathrsfs}

\begin{document}

\title{Numerical spectral synthesis of breather gas for the focusing nonlinear Schr\"odinger equation}

\author{Giacomo Roberti}
\author{Gennady  El} 
\affiliation{Department of Mathematics, Physics and Electrical
Engineering, Northumbria University, Newcastle upon Tyne, NE1 8ST, United
Kingdom}
\author{Alexander Tovbis}
\affiliation{Department of Mathematics, University of Central Florida, Orlando, USA}
\author{Fran\c{c}ois Copie}
\author{Pierre Suret}
\author{St\'ephane Randoux}
\email{stephane.randoux@univ-lille.fr}
\affiliation{Univ. Lille, CNRS, UMR 8523 - PhLAM -
  Physique des Lasers Atomes et Mol\'ecules, F-59 000 Lille, France}

\begin{abstract}
  We  numerically realize breather gas for the focusing nonlinear Schr\"odinger equation. This is done by building  a random ensemble of $N \sim 50$ breathers  via the Darboux transform recursive scheme in high precision arithmetics. Three types of breather gases are synthesized according to the three prototypical spectral configurations  corresponding  the Akhmediev, Kuznetsov-Ma and Peregrine breathers as elementary quasi-particles of the respective gases.   The interaction properties of the constructed breather gases are investigated by propagating through them a `trial' generic breather (Tajiri-Watanabe) and comparing the mean propagation velocity  with the  predictions of the recently developed spectral kinetic theory (El and Tovbis, PRE 2020).
\end{abstract}

\maketitle

\section{Introduction} 
\label{sec:intro}

The study of nonlinear random waves in physical systems well described at leading order by the so-called integrable equations, such the Korteweg-de Vries (KdV) or nonlinear Schr\"odinger (NLS) equations has recently become the topic of intense research in several areas of nonlinear physics, notably in oceanography and nonlinear optics. This interest is motivated by the complexity of  
many natural or experimentally observed nonlinear wave phenomena often requiring a statistical description even though the underlying physical model is, in principle, amenable to the well-established mathematical techniques of integrable systems theory such as inverse scattering transform (IST) or finite-gap theory (FGT) \cite{Novikov_book}.  An intriguing interplay between integrability and randomness in such systems is nowadays associated with the concept of {\it integrable turbulence} introduced by V. Zakharov in  \cite{Zakharov:09}. The integrable turbulence framework is particularly pertinent to the description of modulationally unstable systems whose solutions, under the effect of random noise, can exhibit highly complex spatiotemporal dynamics that are adequately described  in terms of turbulence theory concepts, such as  the distribution functions, ensemble averages, correlations etc.

Solitons and breathers are the elementary ``quasiparticles'' of nonlinear wave fields in integrable systems which can form ordered coherent structures such as modulated soliton trains  and  dispersive shock waves \cite{maiden_solitonic_2018, el_dispersive_2016}, ``superregular  breathers'' \cite{Gelash:14, Kibler:15} or ``breather molecules'' \cite{Xu:19}.  Furthermore, solitons and breathers  can form {\it irregular} structures or statistical ensembles that can be viewed as soliton and breather gases. The nonlinear wavefield in such integrable gases represents a particular case of integrable turbulence \cite{Zakharov:09,Soto:16,Akhmediev:16,Randoux:14,Walczak:15,Suret:16,Michel:20}. The observations of soliton and breather gases in the ocean have been reported in \cite{Costa:14, Wang:18, osborne_highly_2019}. Recent laboratory experiments on the generation of shallow-water and deep water soliton gases were reported in \cite{Redor:19} and \cite{Suret:20} respectively. It has also been demonstrated  that the soliton gas dynamics in the focusing NLS equation provides a remarkably good description of the statistical properties of the nonlinear stage of spontaneous modulational instability \cite{Gelash:19}.

Analytical description of soliton gases was initiated by Zakharov in ref. \cite{Zakharov:71}, where a spectral kinetic equation for KdV solitons was derived using an IST based phenomenological procedure of computing an effective adjustment to a soliton's  velocity in a {\it rarefied} gas due to its collisions with other solitons, accompanied by appropriate phase-shifts. Zakharov's kinetic equation for KdV soliton has been generalized to the case of a {\it dense gas} in ref \cite
{el_thermodynamic_2003} using  the spectral finite-gap theory. Within this theory, a uniform (equilibrium) soliton gas is modelled by a special infinite-phase, thermodynamic type limit of finite-gap KdV solutions. The   kinetic description of the non-equilibrium soliton gas is then enabled by considering the same thermodynamic limit for the associated modulation (Whitham) equations. The resulting kinetic equation describes the evolution of the density of states (DOS) defined as the density function in the spectral (IST) phase plane of soliton gas. The spectral construction of the KdV soliton gas in ref. \cite{el_thermodynamic_2003} has been generalized to the soliton gas of the focusing NLS equation in \cite{GEl:05, GEl:20}. The latter work \cite{GEl:20} provides also the spectral kinetic description of a breather gas (BG), which is the main subject of the present work. 

An isolated generic breather can be broadly viewed as a soliton on the plane wave (or finite) background. The 1D-NLSE equation supports a large family of breather solutions that have attracted a particular interest due to their explicit analytic nature and the potential for modeling the rogue wave events in the ocean and in nonlinear optical fibers \cite{Akhmediev:09b,Akhmediev:09c,Kedziora:12,Genty:10,Dudley:19}. Three types of breathers, namely the Akhmediev breather (AB), the Kuznetsov-Ma (KM) breather and the Peregrine soliton (PS) have aroused significant research interest, see \cite{Kuznetsov:77,Peregrine:83,Akhmediev:86,Kibler:10,Chabchoub:11,Randoux:16a} and references therein. AB, KM breather and PS represent special cases of a generic breather called the Tajiri-Watanabe (TW) breather \cite{Tajiri:98}. A simplest example of breather gas can be viewed as an infinite random ensemble of the TW breathers \cite{GEl:20}. By manipulating the spectral parameters the TW breather gas can be reduced to the AM, KM and PS gases as well as to the fundamental soliton gas. The latter is achieved by vanishing the plane wave background of the TW breather gas \cite{GEl:20}.

The present paper has two goals: (i) numerical realization of a breather gas; (ii) verification of the spectral theory of breather gas developed in \cite{GEl:20}. 

Numerical realization of a breather gas as a large ensemble of TW breathers with prescribed parameters represents a challenging problem. Numerical methods for the construction of breather solutions of the 1D-NLSE suffer from accuracy problems that prevent the numerical synthesis of breathers of order $N \gtrsim 5$ \cite{Akhmediev:88,Kedziora:13}. In the context of soliton gases this latter difficulty has been recently resolved by Gelash and Agafontsev \cite{Gelash:18} via the application of the so-called dressing method combined with  high precision  numerical computations. 
In this paper, we extend  the  algorithm of \cite{Gelash:18} to numerically realize various breather gases and verify some predictions of the spectral kinetic theory of \cite{GEl:20}. In particular we demonstrate that random ensembles of $N \sim 50$ breathers can be build via the Darboux transform recursive scheme in high precision arithmetics. To our knowledge, this represents an improvement of an order of magnitude compared to the results reported in previous numerical works. In addition we show that the construction method can be used to provide evidence of the space-time evolution of the generated breather gases. This feature cannot be achieved by using direct numerical simulations of the 1D-NLSE due to the inevitable presence of modulational instability that quickly desintegrates the plane wave background.

The paper is organized as follows. In Section~\ref{sec:synthesis} we present the algorithm of the spectral synthesis of breather gas
using the Darboux transform. This algorithm is then realized numerically using the high precision arithmetics.  In Section~\ref{sec:breather} we numerically study the interactions in breather gases and compare the results of the numerical simulations with the theoretical predictions of the breather gas kinetic theory of Ref~\cite{GEl:20}. Specifically, we consider the propagation of the `trial' breather  through a homogeneous breather gas for three prototypical configurations: Akhmediev, Kuznetsov-Ma and Peregrine gases. The study of interaction in the gas of Akhmediev breathers has revealed some special features that have required further development of the theory of Ref~\cite{GEl:20}. The Appendix provides the identification of the interaction kernel in the breather gas with the  position shift formula in two-breather collisions, obtained in earlier works.

\section{Nonlinear spectral synthesis of breather gases}\label{sec:th_framework}
\label{sec:synthesis}

\subsection{Soliton and breather ensembles in the 1D-NLSE: an overview}

We consider the integrable one-dimensional focusing NLS equation (1D-NLSE) in the following form: 
\begin{equation}\label{nlse}
  i \psi_t +  \psi_{xx} +  2 \,  |\psi|^2 \, \psi=0,
\end{equation}
where $\psi(x,t)$ represents the complex envelope of the wave field
that evolves in space $x$ and time $t$.

In the inverse scattering transform (IST) method, the 1D-NLSE (\ref{nlse}) is
represented as a compatibility condition of two linear equations \cite{Zakharov:72,Novikov_book},
\begin{equation}\label{LP1}
\Phi_x=
 \begin{pmatrix}
-i \lambda & \psi \\ -\psi^* & i \lambda \\
 \end{pmatrix}
 \Phi, 
\end{equation}
\begin{equation}\label{LP2}
\Phi_t=
 \begin{pmatrix}
-2 i \lambda^2+i|\psi|^2 & i \psi_x +2\lambda \psi \\ i \psi_x^* - 2\lambda \psi^*
& 2 i \lambda^2 -i|\psi|^2 \\
\end{pmatrix} 
\Phi,
\end{equation}
where $\lambda$ is a (time-independent) complex spectral parameter
and $\Phi(x,t,\lambda)=(r(x,t,\lambda),s(x,t,\lambda))^T$ is a
column vector. The spatial linear operator (\ref{LP1}) and the temporal
linear operator (\ref{LP2}) form the Lax pair of Eq. (\ref{nlse}).
For a given potential $\psi(x,t)$ the problem of finding the scattering data 
$\sigma[\psi]$ (also sometimes called the IST spectrum) and the corresponding
scattering solution $\Phi$ specified by the spatial equation (\ref{LP1}) is called 
the Zakharov-Shabat (ZS) scattering problem \cite{yang2010nonlinear}. 
The ZS scattering problem is formally analogous to calculating the Fourier
coefficients in Fourier theory of linear systems, hence the term `Nonlinear Fourier
Transform' is often used in the context of telecommunications systems research,
particularly in the context  of periodic boundary conditions \cite{Wahls:15,Le:17,Turitsyn:17}.

For spatially localized potentials $\psi$ such that $\psi(x,t) \rightarrow 0$
as $|x| \rightarrow \infty$, the complex eigenvalues $\lambda$ are generally
presented by a finite number of discrete points with $\Im(\lambda) \ne 0$ 
(discrete spectrum) and the real line $\lambda \in \mathbb{R}$ (continuous spectrum).
The scattering data $\sigma(\psi)$ consist of a set of $N$ discrete eigenvalues $\lambda_n$ 
$(n = 1, ..., N)$ , a set of $N$ norming constants $C_n$ for each $\lambda_n$ and the
so-called reflection coefficient $\rho (\xi)$,
\begin{equation}
\sigma(\psi)=\{ \rho (\xi) ; \, \lambda_n, \, C_n \}
\end{equation}  
where $\xi \in \mathbb{R}$ denotes the continuous spectrum component.
In this setting where the wavefield $\psi$ lives on a zero background (ZBG), 
the discrete part of the IST spectrum is related
to the soliton content of the wavefield whereas the continuous
part of the IST spectrum is related to the nonlinear dispersive radiation \cite{yang2010nonlinear}. 

A special class of (reflectionless) solutions of Eq. (\ref{nlse}),
the $N$-soliton solutions (N-SS's), exhibits only a discrete spectrum ($\rho(\xi)=0$)
consisting of $N$ complex-valued eigenvalues $\lambda_n$, $n=1,..., N$ and $N$ associated
complex-valued norming constants. The IST formalism has been extensively applied
to examine the processes of interaction, collision and synchronisation in N-SS's,
see e.g. ref. \cite{yang2010nonlinear,sun:16}.
The numerical synthesis of N-SS's can be 
achieved in standard computer simulations (double precision, $16$-digits) up
to $N \sim 10$ \cite{Gelash:18}. 
On the other hand the numerical synthesis of N-SS's with $N$ large represents
a challenging problem that has been resolved only recently  \cite{Gelash:18}.
Combining the so-called dressing method and numerical calculations made using high numerical
precision (a $100$-digits precision is typically necessary for the synthesis of N-SS's with $N\sim100$),
the numerical synthesis of soliton gases (SGs), i. e. large ensembles of N-SS's characterized by a given spectral distribution,
has been demonstrated in ref. \cite{Gelash:18}. 
The opportunity to synthesize numerically large soliton ensembles has opened the way to the experimental
generation of strongly nonlinear wavefields with a pure solitonic content.
In particular recent experiments made in a one-dimensional water tank with 
deep-water surface gravity waves have revealed that the controlled
synthesis of dense SGs can be achieved in hydrodynamics  \cite{Suret:20}.
Moreover, it has been also recently
shown that the so-called bound state SGs provide a model that describes well  
the nonlinear stage of the noise-induced modulation instability \cite{Gelash:19}. 

In addition to the soliton solutions living on zero background, the focusing NLS equation (\ref{nlse}) admits a large variety of 
 solutions living on a nonzero (plane wave) background. 
The IST theory for the focusing nonlinear Schr\"odinger equation with nonzero boundary conditions (NZBC) at infinity has been reported in ref. \cite{Biondini:14, Biondini:15, Ma:79}. 
As in IST with zero boundary conditions, the scattering data $\sigma[\psi]$ in the IST with NZBC consist
of a set of $N$ discrete complex-valued eigenvalues $\lambda_n$, a set of $N$ associated
norming constants $C_n$ and the reflection coefficient $\rho (\lambda)$.
In IST with NZBC, the continuous spectrum does not live on the real axis $\mathbb{R}$ but
on $\mathbb{R} \cup [-iq_0,i q_0]$ where $q_0>0$ represents the amplitude
of the plane wave background \cite{Biondini:14,Biondini:15}. 

The focusing NLS equation with NZBC possesses a rich family of purely solitonic solutions 
(reflectionless potentials, $\rho (\lambda)=0$) named breathers or sometimes solitons on finite background.
The generic ``elementary'' breather parametrized by one single complex-valued eigenvalue ($N=1$) in
the framework of IST with NZBC is the so-called Tajiri-Watanabe breather \cite{Tajiri:98}. 
This elementary solution reduces under certain limits to the solutions found over
the years by Kuznetsov and Ma \cite{Kuznetsov:77,Ma:79}, Peregrine \cite{Peregrine:83},
and Akhmediev \cite{Akhmediev:86}. Using the dressing method, Zakharov and Gelash
constructed a class of two-soliton solutions on finite background, termed superregular breathers and  corresponding to small initial perturbations
of a constant background \cite{Zakharov:13}. This was generalized to several pairs
of breathers in ref. \cite{Gelash:14,Gelash:18b}. Note that most of these breather
solutions of Eq. (\ref{nlse}) have been experimentally realized in hydrodynamics and in optics
\cite{Kibler:10,Chabchoub:11,Chabchoub:12a,Chabchoub:12b,Frisquet:13,Kibler:12,Kibler:15,Xu:19,Dudley:09,Goossens:19}.

\subsection{Darboux transform-based synthesis of breather gases}\label{Darboux}
\label{sec:darboux}
\begin{figure*}[!t]
  \center
  \includegraphics[width=1.\textwidth]{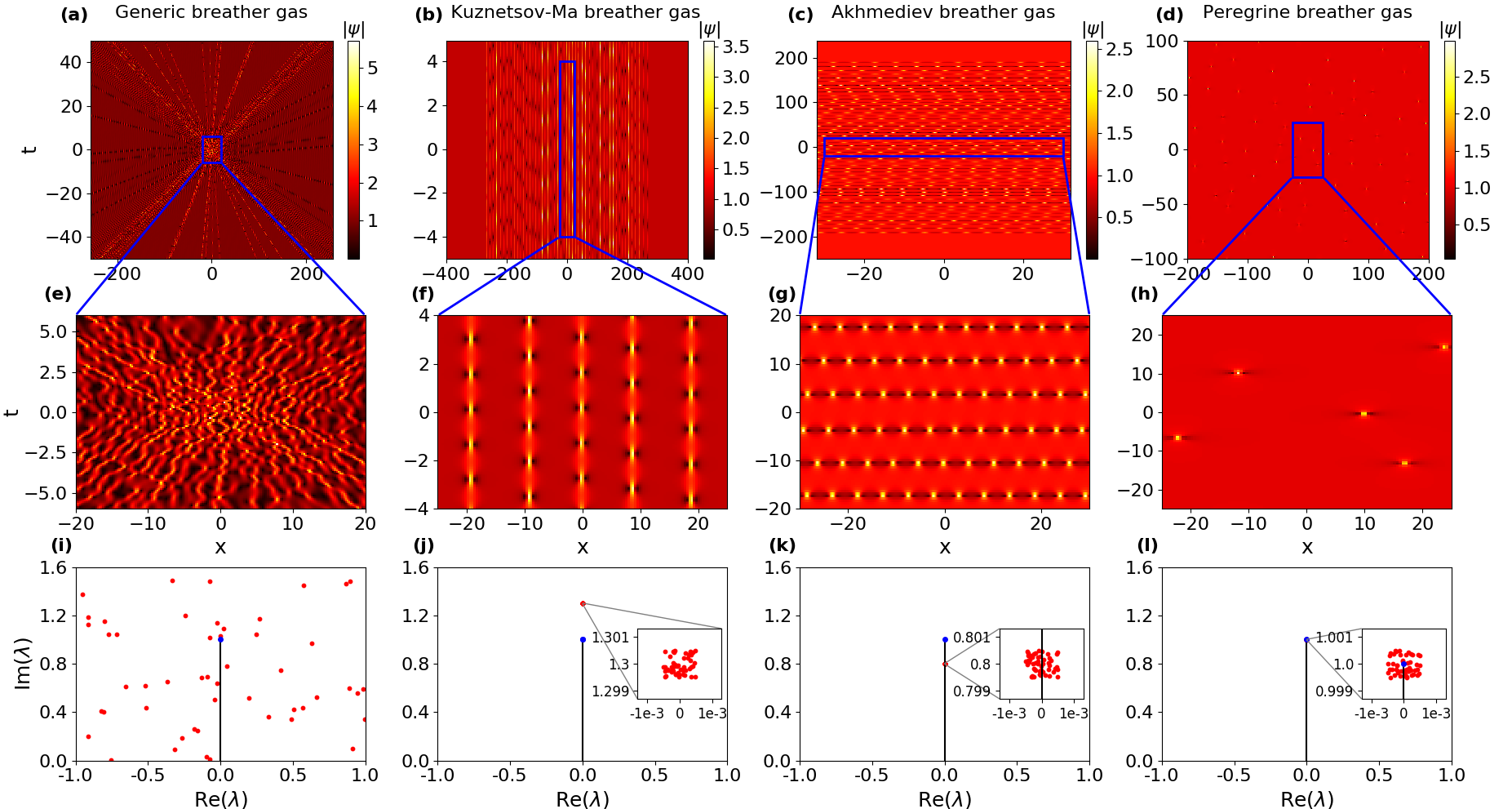}
  \caption{Numerical synthesis of a generic BG (left column (a), (e), (i)) and
    of three single-component BGs: a KM-BG (second column (b), (f), (j)),
    a AB-BG( third columnn (c), (g), (k)) and a PS-BG (fourth column
    (d), (h), (i)). The four BGs are parametrized by $N=50$ complex eigenvalues $\lambda_n$,
    see bottom row.
    The first row (a)--(d) represents the space-time
    evolution of the BGs, with the second row (e)--(h) being an enlarged view of some
    restricted region of the $(x-t)$ plane. The third row (i)--(l) represents the
    spectral portraits of each BG with the vertical line between $0$ and $+i$ being
    the branchcut associated with the plane wave background. Each point in the upper
    complex plane in (i), (j), (k), (l) represents a discrete eigenvalue in the
    IST problem with NZBC. The eigenvalues parametrizing the single-component BGs
    are densely placed in a small square region which is centered around a point
    $\lambda_0$ of the imaginary vertical axis and which is strongly enlarged in the insets
    shown in (j), (k), (l). The $x_j$ are uniformly distributed in the range
    $[-1,1]$ for the generic gas (a) and for the Peregrine gas (d) while they are uniformly distributed
    in the range $[-32,32]$ for the KM gas (b) and the AB gas (c).
  }
\end{figure*}

The recent interest in studying the breather solutions of various kind has been fuelled by the rogue wave
research, see e.g. \cite{Dematteis:19} and references therein. The prototype rogue-wave solutions
represent coherent structures of large amplitude,
strongly localized in both space and time, on an otherwise quiescent background
\cite{Dubard:10,Dubard:11,Gaillard:11,Ankiewicz:11,Ohta:12,Kedziora:12,Kedziora:13,Akhmediev:09b,He:13,Sun:20}.
In this context the Darboux transform has been extensively used as a reliable method
to generate higher-order breather solutions of Eq. (\ref{nlse}), i.e. reflectionless
solutions of the focusing 1D-NLSE with NZBC \cite{Gu:05,Akhmediev:88,Akhmediev:91,Akhmediev:09,Guo:12}.
Note that the Darboux transform
is now also used in the context of nonlinear eigenvalue communication to build
ordered soliton ensembles used to carry out the transmission of information in fiber optics
communication links \cite{Garcia:19,Le:17,Turitsyn:17}. 

The Darboux method is a recursive transformation scheme where a ``seeding solution''
of the focusing 1D-NLSE is used as a building block for the construction of a higher-order
solution through the addition of one discrete eigenvalue.
Here we give a brief review of the Darboux transform method used for the generation of
higher-order breathers. We largely follow the exposition given in ref. \cite{Kedziora:11,Kedziora:13}
but other important references where this method is described and used are
ref. \cite{Gu:05,Akhmediev:88,Akhmediev:91,Akhmediev:09,Guo:12}.

In the IST for the 1D-NLSE  with NZBC, the seeding
solution commonly used at the first step of the recursive process of constructing
a higher-order breather solution is the plane wave
solution of Eq. (\ref{nlse}) with unit amplitude, i.e. $\psi_0(x,t)=e^{2 \, i \, t}$.
The first-order breather (Tajiri-Watanabe) $\psi_1(x,t)$ parametrized by the complex
eigenvalue $\lambda_1$ is obtained by
\begin{equation}\label{Darboux_1}
\psi_1(x,t)=\psi_0(x,t)+\frac {2(\lambda_1^*-\lambda_1)s_{1,1}r_{11}^*}{|r_{1,1}|^2+|s_{1,1}|^2}.
\end{equation}  
The functions $r_{1,1}(x,t)$ and $s_{1,1}(x,t)$ in Eq. (\ref{Darboux_1}) are obtained
by setting $j=1$ in the following expressions 
\begin{equation}\label{r11_s11}
  \begin{split}
    r_{1,j}(x,t)=2i e^{-it} \sin(A_j),
    \\
    s_{1,j}(x,t)=2 e^{it} \cos(B_j),
    \end{split}
\end{equation}  
where $A_j$ and $B_j$ are given by
\begin{equation}\label{A_B}
  \begin{split}
    A_j=\frac{1}{2} \left( \arccos \left( \frac{\kappa_j}{2} \right) + (x-x_j) \kappa_j - \frac{\pi}{2} \right)
    \\
  +  (t - t_j) \kappa_j \lambda_j,     
  \\
  B_j=\frac{1}{2} \left( - \arccos \left( \frac{\kappa_j}{2} \right) + (x-x_j) \kappa_j - \frac{\pi}{2} \right)
  \\
  +  (t - t_j) \kappa_j \lambda_j,       
  \end{split}
\end{equation}  
with $\kappa_j=2\sqrt{1+\lambda_j^2}$. The parameters $(x_j,t_j)$ are connected with the
complex norming constants $C_j$ in the IST with NZBC \cite{Akhmediev:88}. 
The first-order breather $\psi_1(x,t)$ is parametrized by the complex eignevalue $\lambda_1$ and 
by the two real parameters $x_1$ and $t_1$. Once the first-order breather
$\psi_1$ is constructed using Eqs. (\ref{Darboux_1}), (\ref{r11_s11}), (\ref{A_B}),
breather solutions of order $n \ge 2$ can be recursively generated by using 
\begin{equation}\label{Darboux_n}
\psi_n(x,t)=\psi_{n-1}(x,t)+\frac {2(\lambda_n^*-\lambda_n)s_{n,1}r_{n,1}^*}{|r_{n,1}|^2+|s_{n,1}|^2}
\end{equation}  
with
\begin{equation}\label{rnp}
  \begin{split}  
    r_{n,p}=
    [
    (\lambda_{n-1}^* - \lambda_{n-1}) s_{n-1,1}^* r_{n-1,1} s_{n-1,p+1}
    \\
    +(\lambda_{p+n-1} - \lambda_{n-1}) |r_{n-1,1}|^2 r_{n-1,p+1}
    \\
    +(\lambda_{p+n-1} - \lambda_{n-1}^*) |s_{n-1,1}|^2 r_{n-1,p+1}
    ]/
    \\
      (|r_{n-1,1}|^2 + |s_{n-1,1}|^2),
  \end{split}
\end{equation}
\begin{equation}\label{snp}
  \begin{split}  
    s_{n,p}=
    [
    (\lambda_{n-1}^* - \lambda_{n-1}) s_{n-1,1} r_{n-1,1}^* r_{n-1,p+1}
    \\
    +(\lambda_{p+n-1} - \lambda_{n-1}) |s_{n-1,1}|^2 s_{n-1,p+1}
    \\
    +(\lambda_{p+n-1} - \lambda_{n-1}^*) |r_{n-1,1}|^2 s_{n-1,p+1}
    ]/
    \\
      (|r_{n-1,1}|^2 + |s_{n-1,1}|^2).
  \end{split}
\end{equation}

Despite the efficiency of the Darboux method for the construction of high-order breather solutions
of Eq. (\ref{nlse}), its practical implementation in numerics
suffers from the same type of issues as those previously mentioned
for the numerical construction of N-SS's. As noted in ref. \cite{Akhmediev:88,Kedziora:13}, problems
of numerical accuracy may prevent the numerical synthesis of breathers of order
$N \gtrsim 5$. In this paper we show that this limit can be overcome by the
implementation of the same strategy as the one used to build N-SS's with N large \cite{Gelash:18}.
Implementing the Darboux recursive scheme in high precision arithmetics using the BOOTS C++ Multiple precision Library, we show that breather solutions of Eq. (\ref{nlse}) can be synthesized up to 
order $N \sim 50$. As will be shown in detail in Sec. \ref{sec:breather}, this provides a numerical tool
that enables one to verify the results of the spectral theory of breather gases recently developed
in ref. \cite{GEl:20}.

Fig. 1(a) shows the space-time evolution of a generic BG, i.e. a breather solution of
Eq. (\ref{nlse}) of order $N=50$ with random spectral charateristics. The amplitude
of the plane wave background is unity ($q_0=|\psi_0|=1$) and the $50$
complex-valued eigenvalues $\lambda_j$ ($j=1-50$) parametrizing the BG are randomly
distributed within some rectangular region of the upper complex plane, see Fig. 1(i).
The parameters $t_j$ are fixed to zero ($t_j=0 \quad \forall j$) and the
randomness of the gas is achieved by uniformly distributing
the $x_j$ in some interval centered around $x_0=0$. Note that the vertical
line between $0$ and $+i$ in Fig. 1(i) represents the so-called branchcut associated
with the plane wave background in the IST formalism of the 1D-NLSE with NZBG,
see e.g. \cite{Biondini:14,Biondini:15,Randoux:16a,GEl:20}. Fig. 1(a) reveals that the
space-time dynamics of the generic BG synthesized in numerical
simulations is highly complicated. In particular, breathers cannot be individualized
due to their strong overlap and interaction. Note also that the maximum amplitude
reached locally in space and time by the incoherent breather ensemble of Fig. 1(a)
does not exceed $\sim 5.5$,
which demonstrates that the multiple breathers are far from a synchronization state
that would eventually produce isolated rogue waves of large amplitude \cite{Bertola:17,Yang:21}. 

We emphasize that BGs shown in the space-time plots of Fig. 1 are not
obtained from a numerical simulation of Eq. (\ref{nlse}).
Taking a BG generated at a given time $t_0$ using the Darboux method
and using this wavefield as initial condition in a numerical simulation of Eq. (\ref{nlse}),
we observe that modulation instability quickly desintegrates the plane wave background
by amplifying the numerical noise inherent to any pseudo-spectral (split-step like) method
commonly used for the numerical integration of the 1D-NLSE. 
On the other hand space-time plots reported in Fig. 1 are obtained 
from a pure spectral (IST) construction based on the Darboux recursive method
which has been implemented in computer simulations made with high numerical precision.
Starting from an ensemble of $N$ complex eigenvalues $\lambda_j$ and $N$
coordinates $(x_j,t_j)$, the BG is synthesized at time $t$ using the Darboux
machinary (Eqs. (\ref{Darboux_1})-(\ref{snp})). 
A $100$ digits precision is typically necessary to synthesize a BG parametrized 
by an ensemble of $N\sim50$ eigenvalues. 
The space-time plots shown in Fig. 1 are obtained by reiterating the same
synthesis at different values of time $t$.

The central concept in the theory of SGs and BGs 
is the density of states (DOS) \cite{lifshits_introduction_1988}
which represents the distribution function $u(\lambda,x,t)$ in the spectral phase space. 
In the context of the 1D-NLSE \eqref{nlse}  the DOS $u(\lambda, x,t)$, where $\lambda=\beta + i \gamma$,
is defined such that $u d\beta d\gamma d x$ is the
number of breather states with complex spectral parameter
$\lambda \in [\beta, \beta+d\beta] \times [\gamma, \gamma +d\gamma]$ contained in a portion of BG within a spatial interval $[x, x+dx]$  at time $t$. 

One-component BGs have been defined in ref. \cite{GEl:20}
as being characterized by a DOS in the form of the Dirac $\delta$ distribution,
i.e. $u(\lambda)= w \, \delta(\lambda - \lambda_0)$ where $w>0$ represents
the mass of the $\delta$ distribution which is centered around one specfic point
$\lambda_0$ in the complex spectral plane. Fig. 1(b-d)(f-h) display the
space-time evolutions together with the spectral portraits (Fig. 1(j-l)) typifying some one-component
BGs of particular interest.

For the Kuznetsov-Ma BG (KM-BG), the
spectral portait consists of the branchcut (associated with the plane wave background
of unity amplitude) and a dense set of $N=50$ spectral points randomly
placed in a small square region of width $\delta=10^{-3}$ centered
around $\lambda_0=1.3 i$, as shown in Fig. 1(j). Fig. 1(b) shows that
the KM-BG is a dense ensemble of individual KM breathers
having all a zero velocity in the $(x,t)$-plane. Contrary to Fig. 1(a) each KM
breather inside the BG can be individualized and it follows the same periodic
time evolution where the time period is fully determined by $\Im(\lambda_0)$.
The randomness in the one-component KM-BG can be seen from the random distance between
individual KM breathers and their random initial phase, see Fig. 1(f). 

The Akhmediev BG (AB-BG) is characterized by the same distribution of the spectrum $\lambda$ as the KM-BG except
that the point $\lambda_0$ around which the multiple discrete eigenvalues are
accumulated is now placed inside the branchcut associated
with the plane wave background, see Fig. 1(k) where $\lambda_0=+0.8i$. As a result, the AB-BG is more naturally characterized by the {\it spectral flux density}, the temporal counterpart of the DOS.
As shown in Fig. 1(c), the AB-BG consists of a random series of individual
ABs having  identical spatial period, which is fully determined
by $\Im(\lambda_0)$. Similarly to the KM-BG, the randomness in the
one-component AB-BG can be seen from the random time separation between
individual Akhmediev breathers and their random relative phases, see
Fig. 1(g).  

It must be mentioned that the density (spatial or temporal) of the AB or KM breather gases cannot be made arbitrary large:
there is a  configuration termed ``breather condensate'' \cite{GEl:20} corresponding to a critically dense breather gas, similar to a soliton condensate numerically realized in \cite{Gelash:19}. 

It is well known that the Peregrine breather can be obtained as the
spatial and temporal infinite period limits of Akhmediev and Kuznetsov-Ma breathers respectively \cite{Akhmediev:09,Kedziora:11}.
In the spectral (IST) domain, the Peregrine breather is obtained by placing
the eigenvalue parametrizing a first-order breather solution of Eq. (\ref{nlse})
exactly at the endpoint $+i$ of the branchcut associated with the plane wave
background of unit amplitude \cite{Randoux:16a}. Following the same approach,
the one-component Peregrine BG (P-BG) is obtained by accumulating a large
number of discrete eigenvalues in a small area surrounding the endpoint of the
branchcut, see Fig. 1(l). As shown in Fig. 1(d) and in Fig. 1(h), the P-BG represents a collection
of individual and identical Peregrine breathers that are randomly positioned
in space and time. 

\begin{figure*}[!t]
  \includegraphics[width=1.\textwidth]{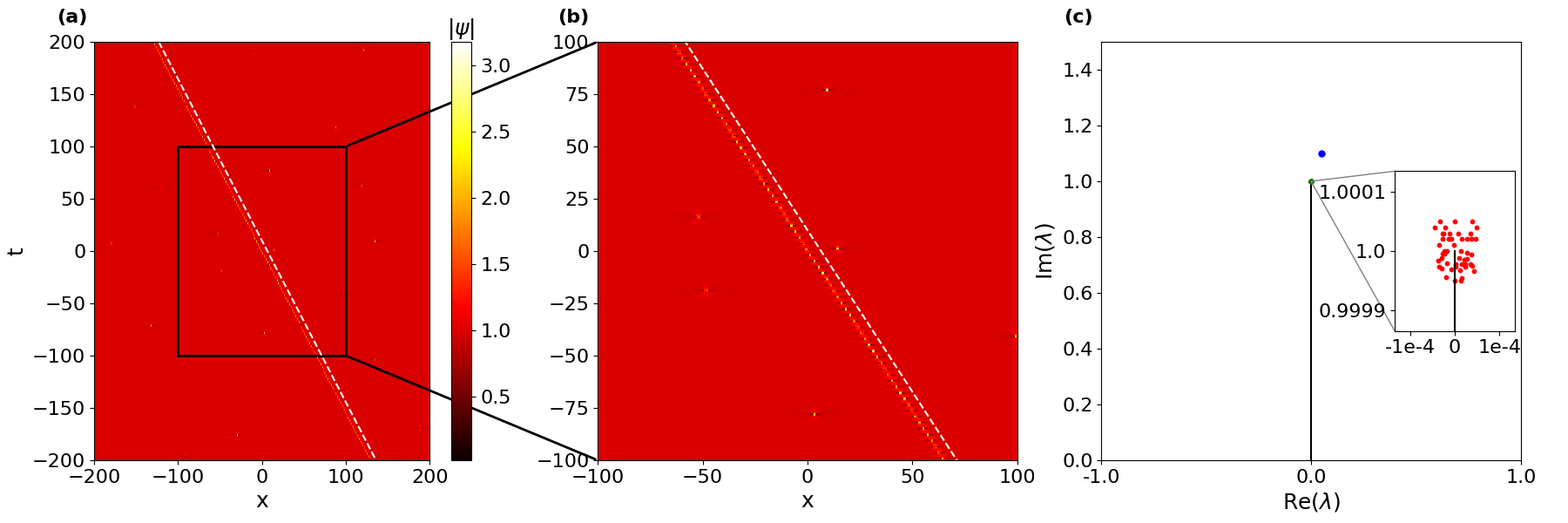}
  \caption{(a), (b) Propagation of a Tajiri-Watanabe breather with the spectral
    parameter $\eta^{[1]}=0.05+1.1i$ inside a
    Peregrine BG. The space-time evolution shown in (b) represents an
    enlarged view of the one shown in (a). The white dashed line in
    (a) and (b) represents the trajectory of the ``free'' Tajiri-Watanabe
    breather propagating on a plane wave background with a group velocity
    given by Eq. (\ref{TW_velocity}). The plot shown in (c) represents the
    spectral portrait associated with the numerical results shown in (a), (b).
    The vertical line between $0$ and $+i$ represents the branchcut associated
    with the plane wave background and the blue point is the discrete eigenvalue $\eta^{[1]}$
    associated with the Tajiri-Watanabe breather propagating in the P-BG. 
    The $50$ spectral points characterizing the P-BG are densely placed around
    $+i$ and they are shown in the inset plotted in (c). 
    }
\end{figure*}

\section{Interactions in Breather gases: comparison between numerical experiments and
  spectral theory}\label{sec:breather}

The analytical theory of BGs has been introduced and developed in ref. \cite{GEl:20}.
It has been shown that spatially non-homogeneous BGs are described by a
kinetic equation formed by a transport equation for
the slowly-varying DOS $u(\lambda,x,t)$ and the integral equation of state relating the gas' velocity to the DOS. In this Section,
we show that some predictions of the spectral theory of BGs can be
verified in simulations involving BGs that have been numerically
synthesized using the methodology described in Sec. \ref{Darboux}. 
In Sec. \ref{spectral_theory}, we provide the key elements of spectral
theory of BGs that are relevant for the comparison between theoretical and
numerical results. In Sec. \ref{numerics_collision}, we examine the
collision between one trial soliton and various single-component BGs.

\subsection{Analytical results from the spectral theory of breather gases}\label{spectral_theory}
\label{sec:spectral_theory}

The nonlinear spectral theory of SGs and BGs for the focusing 1D-NLSE
developed in ref. \cite{GEl:20} provides a full set of equations characterizing the
macroscopic spectral dynamics in a spatially nonhomogeneous BG. 

An important
result of the theory is the so-called equation of state which provides the
mathematical expression of the modification of the mean velocity of a ``tracer'' breather
due to its interaction with other breathers in the gas. 

The group velocity (in the $(x,t)$-plane) of a first-order breather
(TW) parametrized by the complex eigenvalue $\lambda \equiv \eta$ (we shall use in this section this latter notation for the spectral parameter to be consistent with notations of ref. \cite{GEl:20} and previous works on the spectral kinetic theory) is given by 
\begin{equation}\label{TW_velocity}
s_0(\eta)=-2\frac{\Im\left[\eta R_0(\eta)\right]}{\Im\left[R_0(\eta)\right]}
\end{equation}
where $R_0(z)=\sqrt{z^2-\delta_0^2}$ with $\delta_0$ the endpoint of the branchcut
corresponding to the plane wave ($\delta_0=i$ for the plane wave of unit amplitude
considered in all the numerical simulations reported in this paper). It is not difficult to see that, if $\eta \in i \mathbb{R} \setminus [-i, i]$ (KM breather) then $s_0(\eta)=0$, while if $\eta \in (-i,i)$ (AB) then $s_0(\eta)=\pm \infty$ depending on the way the limit $Re (\eta) \to 0$ in \eqref{TW_velocity} is taken (either from the left or right side of the branch cut).

As shown in ref. \cite{GEl:20}, the equation of state of a BG reads
\begin{equation}
s(\eta)=s_0(\eta)+\int _{\Lambda^+}\Delta(\eta,\mu)\big[s(\eta)-s(\mu)\big]u(\mu)|\mathit{d}\mu|
\label{eq:s_int}
\end{equation}
where $\Lambda^+$ is the $2$D compact support of the DOS $u(\eta)$ (defined earlier in Section~\ref{sec:darboux}) located in the upper half plane $\mathbb{C}^+$ of the complex spectral  plane, 
\begin{equation}
  \begin{aligned}
    \Delta(\eta,\mu)=\frac{1}{\Im\left[R_0(\eta)\right]}\Big[\ln\Big|\frac{\mu-\bar{\eta}}{\mu-\eta}\Big|
      \\
      +\ln\Big|\frac{R_0(\eta)R_0(\mu)+\eta\mu-\delta_0^2}{R_0(\bar{\eta})R_0(\mu)+\bar{\eta}\mu-\delta_0^2}\Big|\Big]\, .
    \end{aligned}
\label{eq:phsh}
\end{equation}
The integral term in Eq. (\ref{eq:s_int}) describes the modification of the
`tracer' breather mean velocity in a gas due to its interaction
with other breathers in the gas having a DOS specified by $u$. The spectral value $\eta$ in \eqref{eq:s_int} can be taken outside $\Lambda^+$---in that case formula \eqref{eq:s_int} describes the mean velocity of a ``trial'' or ``test'' TW  breather with the eigenvalue $\eta$ propagating through a breather gas with DOS  supported  $\Lambda^+$.

\begin{figure*}[!t]
  \includegraphics[width=1.\textwidth]{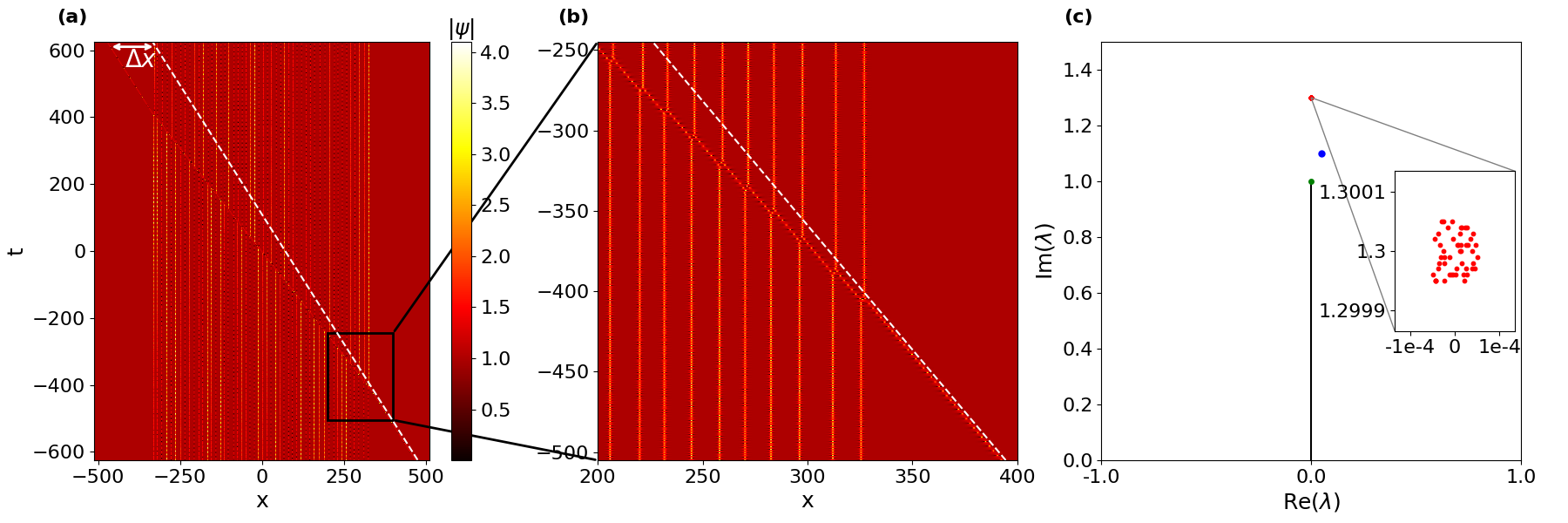}
  \caption{(a), (b) Propagation of a TW breather with the spectral
    parameter $\eta^{[1]}=0.05+1.1i$ inside a
    Kuznetsov-Ma BG. The space-time evolution shown in (b) represents an
    enlarged view of the one shown in (a). The white dashed line in
    (a) and (b) represents the trajectory of the ``free'' TW
    breather propagating on a plane wave background with a group velocity
    given by Eq. (\ref{TW_velocity}). The plot shown in (c) represents the
    spectral portrait associated with the numerical results shown in (a), (b).
    The vertical line between $0$ and $+i$ represents the branchcut associated
    with the plane wave background and the blue point is the discrete eigenvalue $\eta^{[1]}$
    associated with the TW breather propagating in the KM-BG. 
    The $50$ spectral points characterizing the KM-BG are densely placed around
    $\eta^{[2]}=1.3i$ and they are shown in the inset plotted in (c).
  }
\end{figure*}

The interaction
kernel $\Delta(\eta,\mu)$ given by Eq. (\ref{eq:phsh}) describes the position shift arising in a
two-breather interaction. We note that the two-breather interactions have been studied in \cite{Li:18}, \cite{Gelash:18b} using the IST, where  different forms  of the expressions for the position shift were obtained. In the Appendix we demonstrate the equivalence of the kernel $\Delta(\eta, \mu)$ given by  \eqref{eq:phsh} 
to the position shift formula obtained for two-breather collisions in previous works.

For a two-component breather gas, the DOS is a superposition of two Dirac delta-functions
centered at the complex spectral points $\eta^{[j]}$ ($j=1,2$)
\begin{equation}\label{dos_2_components}
  u(\eta)=\sum_{j=1}^2 w^{[j]}\delta(\eta-\eta^{[j]})
\end{equation}  
where $ w^{[j]}$ are the weights of the components.
For the DOS specified by Eq. (\ref{dos_2_components}),
Eq. (\ref{eq:s_int}) yields the following linear system 
 for the gas' component velocities $s^{[j]} \equiv s(\eta^{[j]}) \, (j=1,2)$ 
\begin{equation}
\begin{aligned}
s^{[1]}&=s_0^{[1]}+\frac{\Delta_{1,2} w^{[2]}(s_0^{[1]}-s_0^{[2]})}{1-(\Delta_{1,2} w^{[2]}+\Delta_{2,1} w^{[1]})}\\
s^{[2]}&=s_0^{[2]}-\frac{\Delta_{2,1} w^{[1]}(s_0^{[1]}-s_0^{[2]})}{1-(\Delta_{1,2} w^{[2]}+\Delta_{2,1} w^{[1]})}
\end{aligned}
\label{eq:s_twocomp}
\end{equation}
where $s_0^{[j]} \equiv s_0(\eta^{[j]}) \, (j=1,2)$, $\Delta_{j,k}=\Delta(\eta^{[j]},\eta^{[k]})$.

In the numerical simulations presented in Sec. \ref{numerics_collision},
we will consider an even simpler situation where a single trial breather
parametrized by the eigenvalue $\eta^{[1]}$ interacts with a one-component breather-gas
having its spectral distribution centered in $\eta^{[2]}$. In such a limit $ w^{[1]}\rightarrow 0$ and
Eqs. (\ref{eq:s_twocomp}) reduces to:
\begin{equation}
\begin{aligned}
s^{[1]}&=\frac{s_0^{[1]}-\Delta_{1,2} w^{[2]}s_0^{[2]}}{1-\Delta_{1,2} w^{[2]}}.\\
s^{[2]}&=s_0^{[2]}.
\end{aligned}
\label{eq:s_12}
\end{equation}
The validity of Eqs. (\ref{eq:s_12}) in the context of the  1D-NLSE dynamics \eqref{nlse} will be verified for the P-BG, the KM-BG and the AB-BG in numerical simulations presented in Sec. \ref{numerics_collision}.
As a matter of fact, formula \eqref{eq:s_12}  can be obtained directly from equation \eqref{eq:s_int} by setting $\eta=\eta^{[1] }\notin \Lambda^{+}$ (the trial breather eigenvalue), and using $u(\mu)=w^{[2]}\delta(\mu - \eta^{[2]})$, 
$s(\eta_2)=s_0^{[2]}$ where 
$\eta^{[2]} \in \Lambda^+$.

\subsection{Interactions in one-component breather gases: Comparison between spectral theory and
  numerical simulations}\label{numerics_collision}

In the numerical simulations presented in this Section, a trial TW breather 
with the spectral parameter $\eta=\eta^{[1]}$ is propagated through various single-component
BGs having their DOS defined by $u(\eta)=w^{[2]}\delta(\eta-\eta^{[2]})$. We define
spectral parameter $\eta^{[2]}$  as $\eta^{[2]}=\alpha \, i$ with $\alpha=1$ for the P-BG,
$\alpha > 1$ for the KM-BG, $\alpha < 1$ for the AB-BG.
Similar to Fig. 1, the spectral portait of the considered BGs consists of the branchcut
(associated with the plane wave background of unity amplitude) and a  ``cluster''
of $N=50$ spectral points randomly placed in a small square region of width $\delta=10^{-4}$
centered around $\eta^{[2]}$. The spectral parameter
$\eta^{[1]}$ is chosen in such a way that $\Re({\eta^{[1]})}>0$ which implies that
the free trial TW breather has a negative group velocity in the $(x-t)$ plane,
see Eq. (\ref{TW_velocity}). 

\subsubsection{Interactions in the Peregrine breather gas}\label{numerics_PBG}

Fig. 2 shows a trial Tajiri-Watanabe breather propagating through a P-BG.
We observe that the trial breather passes through the P-BG without change in
its group velocity. This confirms the theoretical result
established in ref. \cite{GEl:20} that the propagation of a trial TW
breather through a P-BG is ballistic. This result can be understood at the
qualitative level by the fact that the interaction cross section between
the trial breather and the individual Peregrine breathers composing the gas
is so weak that the propagation of the trial breather is unaffected by the P-BG. 

\subsubsection{Interactions in the Kuznetsov-Ma breather gas}\label{numerics_KMBG}

Fig. 3 shows a trial TW breather propagating through a KM-BG.
Contrary to Fig. 2, the multiple interactions between the trial breather
and the KM breathers composing the KM-BG now significantly
influences the propagation ot the trial breather, see Fig. 3(a) and 3(b)
for a comparison between the trajectory of the free
Tajiri-Watanabe breather (in white dashed lines) and the trajectory followed
by the trial breather in the KM-BG. As shown in Fig. 3(b), the trial breather
acquires a significant space shift each time that its trajectory
intersects the trajectory of an individual KM breather composing the BG.
At the macroscopic scale, this produces a velocity change of the
trial breather inside the KM-BG. This leads to a spatial shift $\Delta X$ in
the position of the trial breather which is measurable when the trial
breather emerges from the KM-BG, see Fig. 3(a).

For the KM-BG, Eq. (\ref{eq:s_12}) simplifies to 
\begin{equation}
s^{[1]} =\frac{s_0^{[1]}}{1-\Delta_{1,2} w^{[2]}}
\label{eq:kmbg}
\end{equation}
given that $s_0^{[2]}=0$. Eq. (\ref{eq:kmbg}) clearly shows that
the group velocity of the trial Tajiri-Watanabe breather is increased
by a factor $1/(1-\Delta_{1,2} w^{[2]})$ due to the interaction with the
KM-BG. 

Note that the space shift $\Delta X$ acquired by the trial breather
as a result of propagation inside the KM-BG simply represents the product
of the number $N$ of interations (equivalently the number of breathers
in the KM-BG) with the elementary space shift $\Delta_{1,2}$ induced
by each interaction: $\Delta X=N \Delta_{1,2}$. This provides an alternative and
straightforward way to check the validity of 
Eq. (\ref{eq:kmbg}) which gives the group velocity of the trial breather
inside the KM-BG. 

\begin{figure}[h]
  \center
  \includegraphics[width=0.53\textwidth]{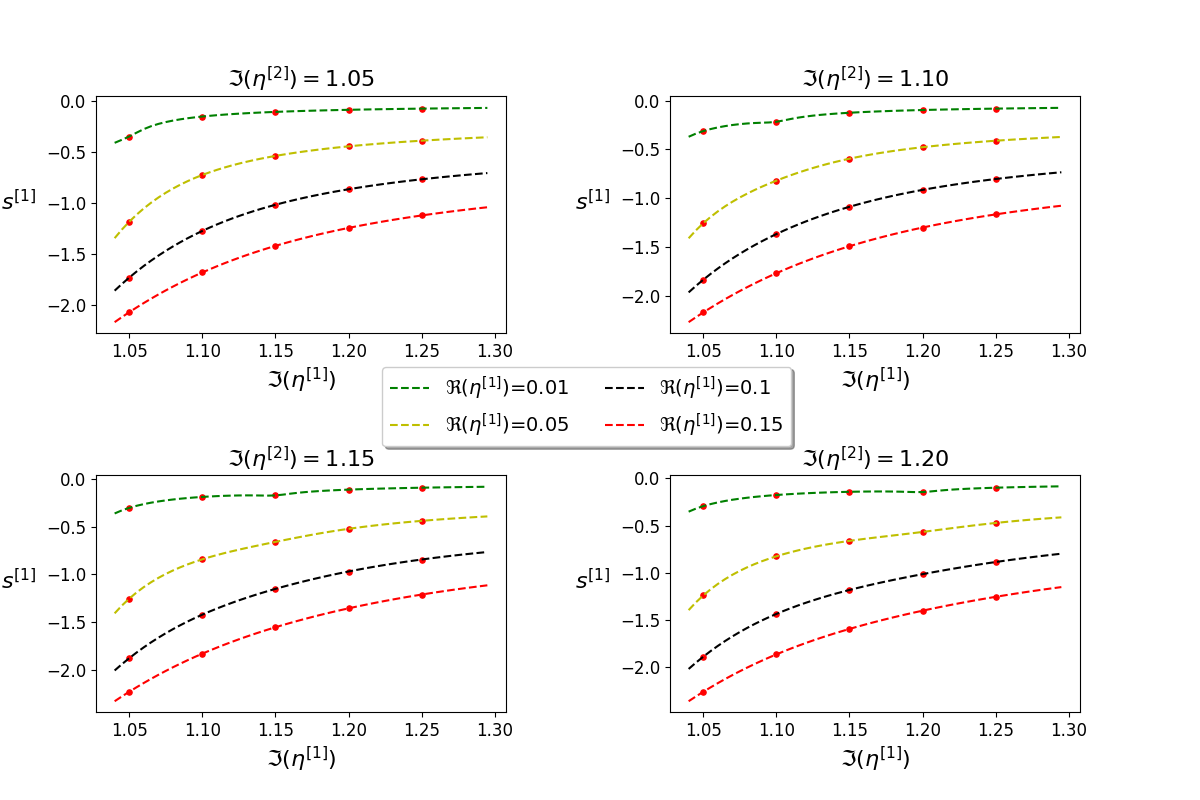}
  \caption{Quantitative verification of the spectral theory of BGs introduced in ref. \cite{GEl:20}.
    Comparison between numerics (red dots) and theory (dashed lines) for the effective
    velocity ($s^{[1]}$) of a trial breather ($\eta^{[1]}$) propagating in a 
    a KM-BG ($\eta^{[2]}$).}
\end{figure}

A set of numerical simulations with different values of the spectral
parameters $\eta^{[1]}$ and $\eta^{[2]}$ has been made to check the validity
of the spectral theory. Different realizations of the KM-BG
have been made and the value of $w^{[2]}$ is determined from numerical
simulations as the ratio between the selected number $N$ of breathers in
the gas over the spatial extension $L$ of the gas: $w^{[2]}=N/L$
As shown in Fig. 4,  we observe full quantitative agreement between
the numerical experiment and the predictions of the spectral theory.

\subsubsection{Interactions in the Akhmediev breather gas}\label{numerics_ABBG}

\begin{figure*}[!t]
  \includegraphics[width=1.\textwidth]{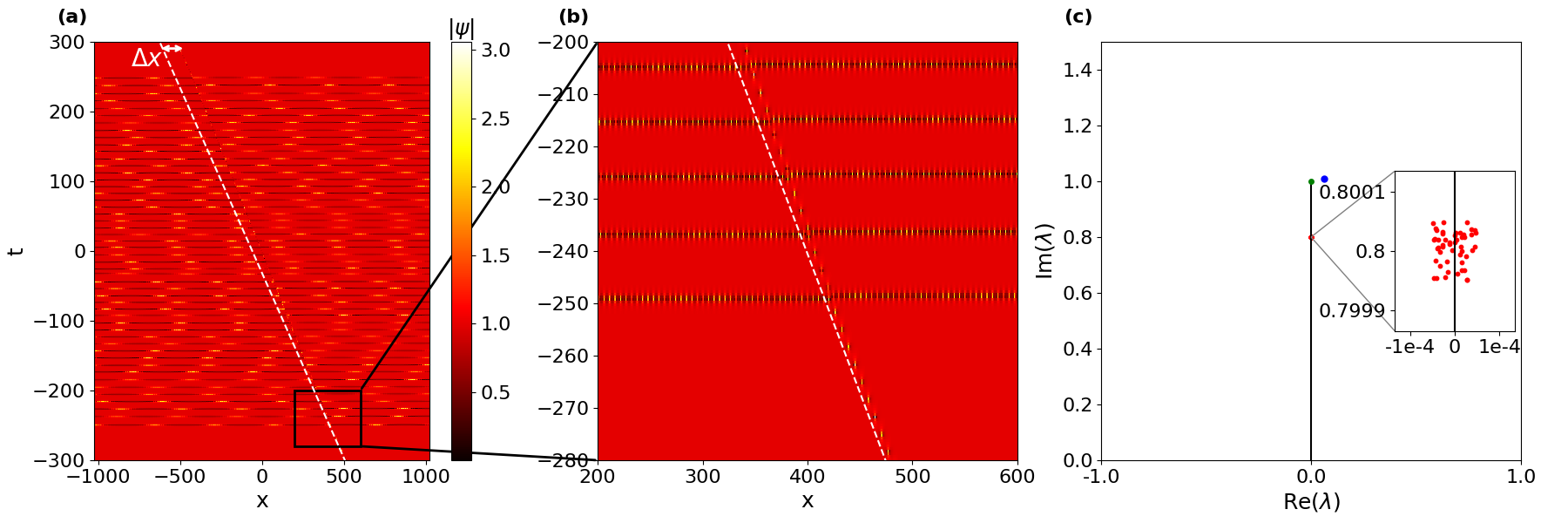}
  \caption{(a), (b) Propagation of a TW breather with the spectral
    parameter $\eta_1=0.06+1.01i$ inside a
    Akhmediev BG. The space-time evolution shown in (b) represents an
    enlarged view of the one shown in (a). The white dashed line in
    (a) and (b) represents the trajectory of the ``free'' TW
    breather propagating on a plane wave background with a group velocity
    given by Eq. (\ref{TW_velocity}). The plot shown in (c) represents the
    spectral portrait associated with the numerical results shown in (a), (b).
    The vertical line between $0$ and $+i$ represents the branch cut associated
    with the plane wave background and the blue point is the discrete eigenvalue $\eta_1$
    associated with the TW breather propagating in the AB-BG. 
    The $50$ spectral points characterizing the KM-BG are densely placed around
    $\eta^{[2]}=0.8i$ and they are shown in the inset plotted in (c).
  }
\end{figure*}

The case of AB-BG is special and requires a separate consideration, particularly because it has not been considered in any detail in \cite{GEl:20}. The AB is a ``static'' object, not localized in space, 
so it is not immediately obvious how to identify the key quantities $u(\eta)$ and $s(\eta)$ for the AB-BG.
 A single AB is a limiting case of the TW breather where the soliton eigenvalue $\eta^{[2]}$  is placed within the branch cut  $[0, i]$ in the upper half plane, The AB-BG is generally characterized by some distribution of soliton eigenvalues along the branch cut. Similar to the  above consideration of KM-BG,  we consider the AB-BG with soliton eigenvalues clustered 
 around a given spectral point $\eta^{[2]}$ (and c.c.) to mimic a one-component gas. 

 As we have already mentioned in Section~\ref{sec:spectral_theory}  the formula \eqref{TW_velocity} for the group velocity of the TW breather implies $|s(\eta)| \to \infty$ 
as $\eta \to \eta^{[2]}$, which is consistent with the delocalized nature of the AB. On the other hand, it can be shown using the results of ref.~\cite{GEl:20}, that in the AB-BG limit the DOS  $u(\eta) \to 0$ while the spectral flux function $v(\eta)=s(\eta) u(\eta)=\mathcal{O}(1)$.  This motivates the following alternative form of the equation of state   \eqref{eq:s_int}:
\begin{equation}\label{vel-correction1}
s(\eta)=s_0(\eta)+\int _{\Lambda^+}\Delta(\eta,\mu)\big[\frac{s(\eta)}{s(\mu)}-1\big]v(\mu)|\mathit{d}\mu|,
\end{equation}
which is more suitable for the characterization of the AB-BG interactions. 
Equation \eqref{vel-correction1} was obtained from \eqref{eq:s_int} 
by substituting $u(\eta)=\frac{v(\eta)}{s(\eta)}$.
Assuming $\Lambda^+$ to be a narrow region surrounding the branch cut $ [0, i]$ and using $|s(\mu)| \gg 1$ for $\mu \in \Lambda^+$
equation 
\eqref{vel-correction1} to leading order becomes
\begin{equation}\label{vel-correction11}
s(\eta)=s_0(\eta)-\int _{\Lambda^+}\Delta(\eta,\mu)v(\mu)|\mathit{d}\mu|.
\end{equation}
Equation \eqref{vel-correction11} describes the modification of the velocity of the TW breather with eigenvalue $\eta$ 
propagating  through the AB-BG characterized by the spectral flux density $v(\mu)$.

An important property of $\Delta(\eta,\mu)$ given by \eqref{eq:phsh} is that 
\begin{equation}\label{prop-Gbr}
 \Delta(\eta,\mu)+ \Delta(\eta,-\bar\mu)=0\quad\text{when}\quad \mu \in[0,i],
\end{equation}
that is, when $\mu$ is on the branch cut  $[0,i]$.
The second variable $\eta$ can take any value in the upper half-plane.
Equation \eqref{prop-Gbr} implies that  $\Delta(\eta,\mu)$ takes  opposite values on the
opposite sides of the branchcut.

 It can further be shown that in the case of a breather gas, whose spectral support $\Lambda^+$ is symmetric with respect to the branch cut  $[0,i]$,
the function $v(\eta)$ also takes opposite values on the opposite sides of $[0,i]$.
Thus the speed of the AB-BG $s(\eta)$ from \eqref{vel-correction1} {\it does not  depend}
on which side of the upper part of the branch cut  $[0,i]$ the domain $\Lambda^+$ or its parts are situated.

Let us now consider a one-component AB-BG with the spectral flux $v(\eta)=w^t\delta(\eta-\eta^{[2]})$, where $\eta^{[2]}\in[0,i]$ and
$w^t$ is a real constant weight.  
As a result, equation \eqref{vel-correction11} assumes a simple form
\begin{equation}\label{vel-correction2}
s(\eta)=s_0(\eta)-w^t\Delta(\eta,\eta^{[2]}),
\end{equation}
We note that the sign of $w^t$, as was explained above, depends on the side of 
$[0,i]$ but the sign of the product $w^t\Delta$ does not. Hence we have
 the general result $s(\eta)-s_0(\eta) <0$ for the propagation of a trial breather through a AB-BG.

We note that formula \eqref{vel-correction2} can be obtained directly from the basic result \eqref{eq:s_12} by using $w^{[2]} \to 0$ and
introducing $w^{[2]} s_0^{[2]}\equiv w^t $. This simple formal consideration, however, does not provide the important information about the sign of $w^{t}\Delta$.

Fig. 5 shows a trial TW breather propagating through a AB-BG.
Similar to Fig. 3, the propagation of the trial breather
is significantly influenced by the the multiple interactions with 
the AB breathers composing the AB-BG, see Fig. 5(a) and 5(b). 
One can see that, in contrast  to the interaction of the trial TW breather  with the KM-BG,
the group velocity of the trial TW breather is reduced in the interaction with the AB-BG, in agreement with Eq.~\eqref{vel-correction2}. 
Indeed, the space shifts observed in Fig. 3(a) and in Fig. 5(a) have opposite signs.  

Similar to the KM-BG interactions, a set of numerical simulations with different values of the spectral
parameters $\eta^{[1]}$ and $\eta^{[2]}$ has been made to check the validity
of equation \eqref{vel-correction2}. Different realizations of the AB-BG
have been produced and the value of $w^{t}$ was determined from numerical
simulations as the ratio between the selected number $N$ of AB  in
the gas over the temporal extension $T$ of the gas: $w^{t}=N/T$
As shown in Fig.~6,  we observe full quantitative agreement between
the numerical experiment and the predictions of the spectral theory.

\begin{figure}[h]
  \center
  \includegraphics[width=0.53\textwidth]{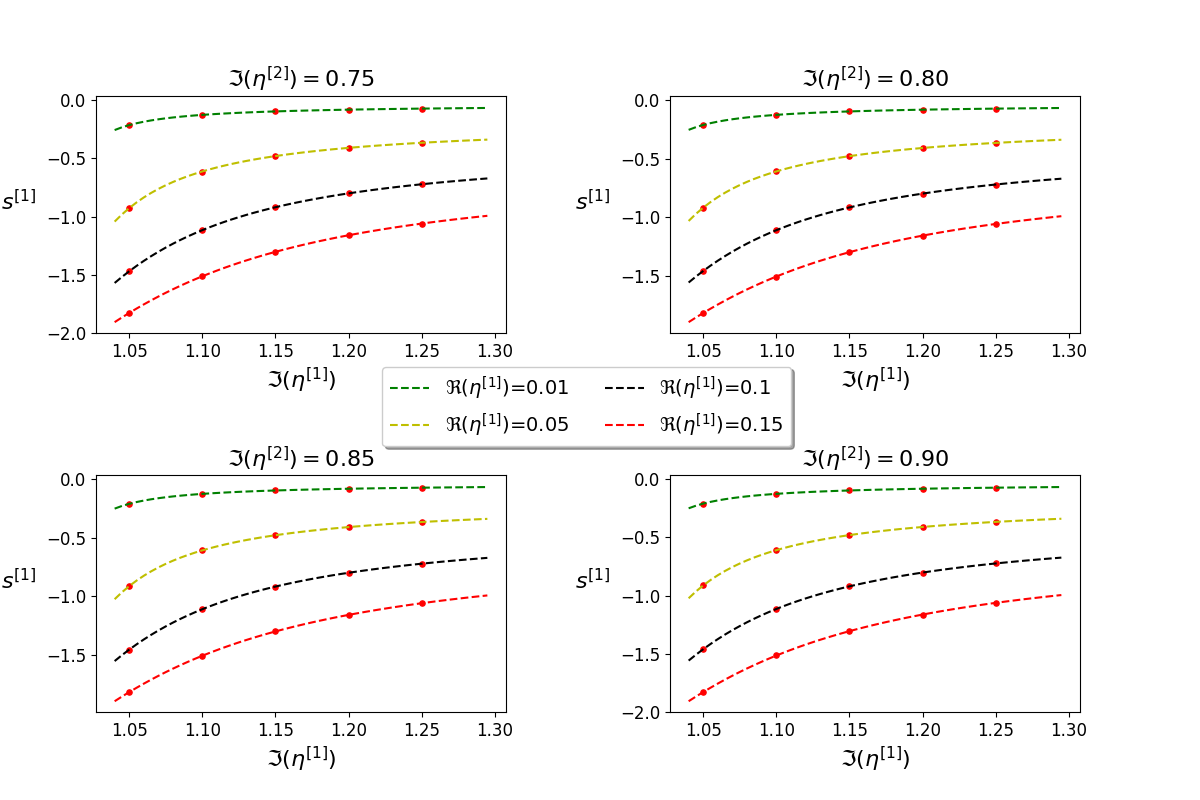}
  \caption{
    Comparison between numerics (red dots) and theory (dashed lines) for the effective
    velocity ($s^{[1]}$) of a trial TW breather ($\eta^{[1]}$) propagating in a 
    a AB-BG ($\eta^{[2]}$).}
\end{figure}

\section{Conclusions}
\label{sec:conclusions}

We have developed a numerical algorithm of  the IST spectral synthesis of breather gases for the focusing 1D-NLS equation. The algorithm is based 
on the recursive Darboux transform scheme realized in high precision arithmetics. Using this algorithm we have synthesized numerically three types of ``prototypical'' breather gases: the Akhmediev, Kuznetsov-Ma and Peregrine gas. 

Using the developed spectral algorithm, the interaction properties of breather gases, predicted by the kinetic theory of ref. \cite{GEl:20} have been tested by propagating through them a `trial' generic TW breather whose effective velocity is strongly affected by the interaction with the gas.  In all cases the theoretically predicted effective  mean velocity of the trial breather propagating through a breather gas demonstrates  excellent agreement with the results of  the numerical simulations. The verification of the theory, despite the inevitable effects of modulational instability present in the 1D-NLSE dynamics, has been made possible due to the whole numerical algorithm being based on the  spectral construction rather than direct simulations of the 1D-NLSE equation.

The quantitative verification of the kinetic theory of breather gases undertaken in this paper is an important step towards a better understanding of this type of a turbulent motion in integrable systems. We also believe that the ability to synthesize numerically BGs represents a step of importance towards the controlled laboratory generation of BGs, possibly following an approach similar to the one recently reported for hydrodynamic SGs \cite{Suret:20}. Finally the possibility to generate numerically breather solutions of order $N \gtrsim 10$ paves the way for further works devoted to the investigation of the properties of localization in space and time of breather solutions of the 1D-NLSE of very high order \cite{Kedziora:13,Gelash:18b,Yang:21}.

\begin{acknowledgments}
This work has been partially supported by the Agence
Nationale de la Recherche through the I-SITE ULNE
(ANR-16-IDEX-0004), the LABEX CEMPI (ANR-11-
LABX-0007) and the Equipex Flux (ANR-11-EQPX-
0017), as well as by the Ministry of Higher Education
and Research, Hauts de France council and European
Regional Development Fund (ERDF) through the CPER
project Photonics for Society (P4S), EPSRC grant (UK)
EP/R00515X/2 (GE), NSF (USA) grant DMS 2009647 (AT) and Dstl (UK) grant DSTLX-1000116851 (GR, GE, SR).
GE, AT and GR thank the PhLAM laboratory at the University of Lille for
hospitality and partial financial support. 
\end{acknowledgments}

\section*{Appendix: Position shift in two-breather interactions}

The two-breather interactions have been studied in refs.~\cite{Li:18}, \cite{Gelash:18b} where the expressions for the phase and position shifts in the interaction of two Tajiri-Watanabe breathers have been derived using the IST analysis. In Section~\ref{sec:spectral_theory} of this paper the interaction kernel in the equation of state \eqref{eq:s_int} for breather gas has been obtained in the form \eqref{eq:phsh}. The natural interpretation of this interaction kernel, consistent with the previously studied cases of KdV and NLS soliton gases, is the  position shift in a two-breather collision. However, the equivalence between formula \eqref{eq:phsh} and the expressions  from \cite{Li:18}, \cite{Gelash:18b} is far from being obvious. Here we establish this equivalence enabling one to extend the phenomenological interpretation of soliton gas kinetics \cite{GEl:05} to breather gases.

We consider the position shift expression from \cite{Li:18}
\begin{equation}
\Delta\bar{\xi}_2=-\ln(\xi_0)/(c_{-,2}\cos\alpha_2)=\Delta(\lambda_2,\lambda_1),
\label{eq:phBi}
\end{equation}
where
\begin{equation}
\xi_0=\frac{d_+-2\left(\cos(\alpha_1-\alpha_2)+c_{-,1}c_{-,2}\right)\cos(\alpha_1-\alpha_2)}{d_+-2\left(\cos(\alpha_1+\alpha_2)-c_{-,1}c_{-,2}\right)\cos(\alpha_1+\alpha_2)}
\end{equation}
with 
\begin{equation}
\begin{aligned}
& c_{\pm,j}=z_j\pm q_0^2/z_j \qquad \lambda_j=(\zeta_j-q_0^2/\zeta_j)/2  \\
& d_{\pm,j}=z_j^2\pm q_0^4/z_j^2 \qquad q_0=-i\delta_0 \\
& d_+=d_{+,1}+d_{+,2} \qquad R_0(\lambda_j)=(\zeta_j+q_0^2/\zeta_j)/2\\
& \zeta_j=R_0(\lambda_j)+\lambda_j=iz_je^{i\alpha_j}.
\end{aligned}
\label{eq:map}
\end{equation}
One can verify that substituting \eqref{eq:map} in \eqref{eq:phsh} and invoking the identities
\begin{equation}
\begin{aligned}
&|\lambda_i|^2=\left(d_{+,i}+2q_0^2\cos\alpha_i\right)/4 \\ 
&d_+=\left(z_1z_2+\frac{q_0^4}{ z_1z_2}\right)\left(\frac{z_1}{z_2}+\frac{z_2}{z_1}\right) \\
&\left(\cos 2 \alpha_1+\cos 2 \alpha_2\right)/2=\cos(\alpha_1+\alpha_2)\cos(\alpha_1-\alpha_2)
\end{aligned}
\end{equation}
yields the phase shift expression \eqref{eq:phBi}.

%

\end{document}